\documentstyle[amstex,preprint,aps]{revtex}

\begin{document}
\title{Non separable Werner states in spontaneous parametric down-conversion}
\author{Marco Caminati,\ Francesco De Martini, Riccardo Perris, Fabio Sciarrino and
Veronica Secondi}
\address{Dipartimento di Fisica and Istituto Nazionale per la Fisica della Materia,\\
Universit\'{a} ''La Sapienza'', Roma 00185, Italy}
\maketitle

\begin{abstract}
The multiphoton states generated by high-gain spontaneous parametric
down-conversion (SPDC) in presence of \ large losses are investigated
theoretically and experimentally. The explicit form for the two-photon
output state has been found to exhibit a Werner structure very resilient to
losses for any value of \ the gain parameter, $g$. The theoretical results
are found in agreement with the experimental data. The last ones are
obtained by quantum tomography of \ the state generated by a high-gain SPDC.
\end{abstract}


\section{Introduction}

Entanglement, the non classical correlation between separated quantum
systems, represents the physical resource lying at the very basis of quantum
information (QI), quantum computation and quantum communication. The current
proliferation of relevant applications of quantum entanglement, ranging from
the one-way quantum computation \cite{Raus01} to the emerging fields of
quantum metrology, lithography, etc. \cite{Boto00} strengthens the need of
new flexible and reliable techniques to generate entangled states with
increasing dimension.

Entangled photonic qubit pairs generated under suitable phase-matching
conditions by the spontaneous parametric down conversion (SPDC) process in a
nonlinear (NL) crystal have been central to many applications ranging from
teleportation \cite{Bosc97} to various quantum key distribution protocols 
\cite{Gisi02}. Today the realization of reliable SPDC sources with large
efficiency, i.e. with large ''brilliance'', able to generate entangled pure
states is of key interest as they largely determine the range of
applications of the sophisticated optical methods required by modern QI.
Recently, high brilliance sources have been realized by using bulk \cite
{Kwia94,Kwia99,Cine04} and periodically poled \cite{Fior04} NL crystals, and
fiber coupled SPDC radiation \cite{Kurt01}. The extension of the attained
results to higher dimensional QI qubit carriers could lead to new
information processing tasks \cite{Durk03} and requires the development of
appropriate technological and theoretical tools. Significant results in this
field are the SPDC\ generation of \ four-photon entangled states with or
without ''{\it quantum-injection}'' \cite
{DeMa2000,Lama01,Pell03,DeMa04,Eibl03,Zhao04}. At the same time, the
sophisticated level attained by QI processing of photonic qubits has
directed a consistent part of the experimental endeavour towards the
solution of more technical problems. For instance, since the losses due to
the photon transmission represent a restriction to the realization of
several QI protocols, classes of entangled states robust against losses have
been investigated and a few experimental schemes to generate such states
have been proposed \cite{Dur00}. In the case of three qubits, entangled
states can be classified as ''GHZ-type'' or as ''W-type'', the latter
exhibiting a larger robustness in bipartite entanglement \cite{Acin01}. The
W-states can be generalized to an arbitrary number of qubits and can be
generated efficiently by spontaneous or stimulated emission in a high gain $%
(HG)$ regime, i.e. when the parametric ''gain'' is large: $g\gg 0$. The
first experimental realization of 3 photons W-states by spontaneous emission
have been recently reported \cite{Eibl04,Walt05b}.

In\ this framework the investigation of the multiphoton state generated in $%
HG$ condition is of fundamental importance, both on a conceptual and
practical sides, e.g. for non-locality test \cite{Reid02,Simo03} or for QI
applications. Recently multiphoton entangled states created by $HG$ SPDC
have been observed \cite{Eise04}. The aim of the present paper is to
investigate relevant aspects of the output wavefunction of the $HG$ scheme.
In Section II, the density matrix representing the 2-photon reduced state
arising from $HG$ SPDC after propagation over lossy channels is analyzed
theoretically. We find that, for any value of $g$, the resulting 2-photon
state is a ''Werner state'' \cite{Wern89}, i.e., a superposition of a
maximally entangled (singlet) state with a fully mixed state. This condition
is investigated thoroughly by exploiting the extensive knowledge available
on the Werner states and modern techniques like the ''entanglement witness''
and the various forms of state entropy \cite{Guhn02,Barb03}. Resilience of
entanglement is theoretically demonstrated for any value of $g$\ in the high
loss $(HL)$\ approximation \cite{Durk04}.\ In Section III, the results of
the theory are compared with the corresponding experimental data obtained by
conventional quantum state tomography $(QST)$.

\section{SPDC in high losses regime: generation of\ Werner states}

In this Section, we theoretically analyze the effect of losses on $HG$ SPDC\
multiphoton state. Recently Durkin {\it et al.} \cite{Durk04} demonstrated
the persistence of some kind of symmetries implying entanglement in
multiphoton SPDC states in presence of polarization independent photon
losses. Here we explicitly derive the expression of the SPDC density matrix
in regime of induced high photon losses through coincidence measurements and
demonstrate that it corresponds to a Werner state for any value of $g$. In
the present model effects of both losses and imperfect detections on the
output states are simulated, as usual, by the insertion of beam splitters on
the two propagation modes ${\bf k}_{i}$ $(i=1,2)$ (Fig.1). The results of
this study demonstrate the resilience of bipartite entanglement for any
value of $g$. This implies that, even in absence of these induced losses,
the initial SPDC\ state is entangled for any $g$, because of the basic
impossibility of creating or enhancing the entanglement by means of local
operations acting on non entangled state \cite{Vedr98}. The\ main
motivations for the present investigation resides in the experimental
entanglement assessment on multi-photon HG fields by introducing light
absorbing filters on the correlated photons paths. The approximate
expression of the density matrix also provides an intuitive explanation of
the behavior of SPDC states in the HG\ and HL regimes.

The Hamiltonian of the SPDC process in the interaction picture reads \cite
{Pell03,DeMa05} 
\begin{equation}
\hat{H}=i\kappa (\hat{a}_{1H}^{\dagger }\hat{a}_{2V}^{\dagger }-\hat{a}%
_{1V}^{\dagger }\hat{a}_{2H}^{\dagger })+h.c.
\end{equation}
where $\hat{a}_{ij}^{\dagger }$ represents the creation operators associated
with the spatial propagation mode ${\bf k}_{i}$, with polarization $%
j=\{H,V\}.$ $H$ and $V$ stand for horizontal and vertical polarization. $%
\kappa $ is a coupling constant which depends on the crystal nonlinearity
and is proportional to the amplitude of the pump beam. This Hamiltonian
generates a unitary transformation $\hat{U}=e^{-i\frac{\hat{H}t}{\hbar }%
\text{ }}$acting on the input vacuum state $\left| 0\right\rangle
_{1H}\left| 0\right\rangle _{1V}\left| 0\right\rangle _{2H}\left|
0\right\rangle _{2V}=\left| 0,0,0,0\right\rangle $. The output state $\left|
\Psi ^{OUT}\right\rangle =\hat{U}\left| 0,0,0,0\right\rangle $ is easily
obtained in virtue of the disentangling theorem \cite{Coll88}: 
\begin{equation}
\left| \Psi ^{OUT}\right\rangle =\frac{1}{C^{2}}\sum\limits_{n=0}^{\infty }%
\sqrt{n+1}\Gamma ^{2}\left| \psi _{-}^{n}\right\rangle  \label{Singletstate}
\end{equation}
where $\left| \psi _{-}^{n}\right\rangle $ is the $n$-generated pairs term: 
\begin{equation}
\left| \psi _{-}^{n}\right\rangle =\frac{1}{\sqrt{n+1}}\sum%
\limits_{m=0}^{n}(-1)^{m}\left| n-m\right\rangle _{1H}\left| m\right\rangle
_{1V}\left| m\right\rangle _{2H}\left| n-m\right\rangle _{2V}
\end{equation}
and $\Gamma =\tanh g$, $C=\cosh g$. The parameter $g\equiv \kappa t_{int}$
expresses the NL gain of the parametric process, being $t_{int}$ the
interaction time. The average number of photon generated per mode is equal
to $\overline{n}=\sinh ^{2}g.$

Let us first consider the contribution $\left| \psi _{-}^{n}\right\rangle
\left\langle \psi _{-}^{n}\right| $ to the overall density matrix. To
investigate the propagation over a lossy channel, a beam splitter$(BS)$\
with transmittivity $\eta $ for any polarization and spatial mode is assumed
to simulate the effect of channel losses and of detector inefficiencies.
Furthermore, perfect detectors with $\eta _{QE}=1$ measure the output state 
\cite{Loud} (Fig.1). The symmetry of the entangled state after losses is
preserved by assuming $\eta $ to be mode- and polarization- independent. The
contribution $\left| \psi _{-}^{n}\right\rangle $ to the SPDC state is
expressed in terms of the BS operators $\left\{ \hat{a}_{ij-IN}^{\dagger
}\right\} $ associated with the input modes $\{{\bf k}_{i}^{IN}\}$: 
\begin{equation}
\left| \psi _{-}^{n}\right\rangle =\frac{1}{\sqrt{n+1}}\frac{1}{n!}(\hat{a}%
_{1H-IN}^{\dagger }\hat{a}_{2V-IN}^{\dagger }-\hat{a}_{1V-IN}^{\dagger }\hat{%
a}_{2H-IN}^{\dagger })^{n}\left| 0,0,0,0\right\rangle
\end{equation}

The BS's couple the input modes $\{{\bf k}_{i}^{IN}\}$ with the transmitted
modes $\{{\bf k}_{i}^{OUT}\}$ and the reflected modes $\{\widetilde{{\bf k}}%
_{i}^{OUT}\}.$ The output state expression is found by substituting the
operators $\left\{ \hat{a}_{ij-IN}^{\dagger }\right\} $ with their
expressions in term of the operators $\left\{ \widehat{a}_{ij-OUT}^{\dagger
}\right\} $, associated with the output transmitted modes $\{{\bf k}%
_{i}^{OUT}\}$, and the operators$\left\{ \widehat{b}_{ij-OUT}^{\dagger
}\right\} $, associated with the output reflected modes $\{\widetilde{{\bf k}%
}_{i}^{OUT}\}$, through the BS input-output matrix \cite{Loud}: 
\begin{equation}
\left( 
\begin{array}{c}
\widehat{a}_{ij-OUT}^{\dagger }(t) \\ 
\widehat{b}_{ij-OUT}^{\dagger }(t)
\end{array}
\right) =\left( 
\begin{array}{cc}
\sqrt{\eta } & i\sqrt{1-\eta } \\ 
i\sqrt{1-\eta } & \sqrt{\eta }
\end{array}
\right) \left( 
\begin{array}{c}
\hat{a}_{ij-IN}^{\dagger }(t) \\ 
\widehat{b}_{ij-IN}^{\dagger }(t)
\end{array}
\right)  \label{beamsplitter}
\end{equation}
The input state $\left| \psi _{-}^{n}\right\rangle $ evolves into an output
state $\left| \psi _{-}^{n}\right\rangle ^{OUT}$ which is defined over the
four transmitted modes and the four reflected modes.\ This state is
expressed through the Fock states: $\left|
n_{1H},n_{1V},n_{2H},n_{2V}\right\rangle _{a}\otimes \left|
n_{1H},n_{1V},n_{2H},n_{2V}\right\rangle _{b}$ where the first term in the
tensor product represents modes transmitted by the BS and hence detected ($%
\widehat{a}-$modes) while the second term expresses the reflected modes ($%
\widehat{b}-$modes).

The density matrix $\sigma ^{OUT}$= $\left[ \left| \psi
_{-}^{n}\right\rangle \left\langle \psi _{-}^{n}\right| \right] ^{OUT}$ of
the $n$-pair term of the SPDC state can be easily obtained from the previous
expressions: 
\begin{eqnarray}
\left[ \left| \psi _{-}^{n}\right\rangle \left\langle \psi _{-}^{n}\right| %
\right] ^{OUT} &=&\frac{1}{n+1}\left( \frac{1}{n!}\right)
^{2}\sum\nolimits_{k,l_{i}}\sum\nolimits_{h,f_{j}}A^{\ast }(h,\left\{
f_{j}\right\} )A(k,\left\{ l_{i}\right\} )\ast \\
&&\ast \left| l_{1},l_{2},l_{3},l_{4}\right\rangle _{a}\otimes \left|
n-k-l_{1},k-l_{2},k-l_{3},n-k-l_{4}\right\rangle _{b}\ast  \nonumber \\
&&\ast _{a}\left\langle f_{1},f_{2},f_{3},f_{4}\right| \otimes
_{b}\left\langle n-h-f_{1},h-f_{2},h-f_{3},n-h-f_{4}\right|  \nonumber
\end{eqnarray}
with 
\begin{eqnarray}
A(x,\left\{ y_{k}\right\} ) &=&%
{n \choose x}%
{n-x \choose y_{1}}%
{x \choose y_{2}}%
{x \choose y_{3}}%
{n-x \choose y_{4}}%
(-1)^{x}\eta ^{y_{1}+y_{2}+y_{3}+y_{4}}(-i\sqrt{1-\eta }%
)^{2n-y_{1}-y_{2}-y_{3}-y_{4}}\ast \\
&&\ast \sqrt{%
y_{1}!(n-x-y_{1})!y_{2}!(x-y_{2})!y_{3}!(x-y_{3})!y_{4}!(n-x-y_{4})!} 
\nonumber
\end{eqnarray}
and $\sum\nolimits_{x,y_{i}}\equiv
\sum\limits_{x=0}^{n}\sum\limits_{y_{1},y_{4}=0}^{n-x}\sum%
\limits_{y_{3},y_{2}=0}^{x}$. Since we are interested in the reduced density
matrix $\rho _{a}^{n}$ defined over the transmitted modes $\{{\bf k}%
_{i}^{OUT}\}$, $\rho _{a}^{n}$= $Tr_{b}\left[ \left| \psi
_{-}^{n}\right\rangle \left\langle \psi _{-}^{n}\right| \right] ^{OUT}$, we
need to trace $\sigma ^{OUT}$ over the undetected reflected modes. The
result is: 
\begin{eqnarray}
\rho _{a}^{n} &=&\frac{1}{n+1}\left( \frac{1}{n!}\right)
^{2}\sum\nolimits_{k,l_{i}}\sum\nolimits_{h,f_{j}}A^{\ast }(h,\left\{
f_{j}\right\} )A(k,\left\{ l_{i}\right\} )\left|
l_{1},l_{2},l_{3},l_{4}\right\rangle _{aa}\left\langle
f_{1},f_{2},f_{3},f_{4}\right| \\
&&\ast \delta (n-k-l_{1},n-h-f_{1})\delta (k-l_{2},h-f_{2})\delta
(k-l_{3},h-f_{3})\delta (n-k-l_{4},n-h-f_{4})  \nonumber
\end{eqnarray}

The final expression for the $n$-pair contribution to the SPDC density
matrix is: 
\begin{multline}
\rho _{a}^{n}=\frac{1}{n+1}\sum\nolimits_{k,l_{i}}\sum%
\nolimits_{h,f_{j}}(-1)^{k+h}(1-\eta )^{2n}S(h,k,l_{2})S(h,k,l_{3})%
\widetilde{S}(h,k,l_{1})\widetilde{S}(h,k,l_{4})  \nonumber \\
\left| l_{1},l_{2},l_{3},l_{4}\right\rangle \left\langle
k-h+l_{1},h-k+l_{2},h-k+l_{3},k-h+l_{4}\right|  \nonumber
\end{multline}
where $S(h,k,p)=\zeta ^{p}\sqrt{%
{k \choose p}%
{h \choose k-p}%
},$ $\widetilde{S}(h,k,p)=\zeta ^{p}\sqrt{%
{n-k \choose p}%
{n-h \choose k-h+p}%
}$ and $\zeta =\frac{\eta }{1-\eta }$.

Up to now we have considered arbitrary, polarization-symmetric losses. In
the following we make the \ additional assumption of very large losses, i.e.
HL, which greatly simplifies our task. Such approximation is expressed by
the relation $\eta \overline{n}<<1$, $\eta \overline{n}$ being the average
number of photon transmitted by the BS per mode.\ This condition enables us
to take into account only the terms of the sum Eq.7 of order $\leq \eta ^{2}$%
, hence considering only matrix element corresponding to no more than two
photons transmitted. As final step, we assume to detect one photon on each
mode ${\bf k}_{i}^{OUT}$ by a 2-photon coincidence technique. In this way
the vacuum terms affecting one or both vectors ${\bf k}_{i}^{OUT}$ are
dropped. In summary, this coincidence procedure guarantees, by
post-selection, that we are dealing only with matrix elements arising from
the tensor product of the states:$\left\{ \left| 1,0,1,0\right\rangle
,\left| 1,0,0,1\right\rangle ,\left| 0,1,1,0\right\rangle ,\left|
0,1,0,1\right\rangle \right\} $, which correspond to the states $\left\{
\left| H\right\rangle _{1}\left| H\right\rangle _{2},\left| H\right\rangle
_{1}\left| V\right\rangle _{2},\left| V\right\rangle _{1}\left|
H\right\rangle _{2},\left| V\right\rangle _{1}\left| V\right\rangle
_{2}\right\} $. The $n$-pair contribution $\rho _{post}^{n}$ to the SPDC
2-photon density matrix hence reads 
\begin{equation}
\rho _{post}^{n}=\frac{1}{6}n(1-\eta )^{2n}\zeta ^{2}\left( 
\begin{array}{cccc}
(n-1) & 0 & 0 & 0 \\ 
0 & (1+2n) & -(n+2) & 0 \\ 
0 & -(n+2) & (1+2n) & 0 \\ 
0 & 0 & 0 & (n-1)
\end{array}
\right)
\end{equation}
We note that the above density matrix has the form of a Werner state $\rho
_{W}=p\left| \Psi _{-}\right\rangle \left\langle \Psi _{-}\right| +\frac{1-p%
}{4}I$, with $p=\frac{(n+2)}{3n}$ which is a mixture with probability $p$ of
the maximally entangled state$\ \left| \Psi _{-}\right\rangle
=2^{-1/2}(\left| H\right\rangle _{1}\left| V\right\rangle _{2}-\left|
V\right\rangle _{1}\left| H\right\rangle _{2})$ and of the\ maximally
chaotic state $I/4$ being $I$ \ the identity operator on the overall Hilbert
space. These states are commonly adopted in QI, since they model a
decoherence process occurring on a singlet state traveling along an
isotropic noisy channel \cite{Niel00}.

The complete density matrix for the SPDC output state is obtained by
substituting the $\rho _{post}^{n}$ matrices into the expression $\rho
_{th}^{II}=\frac{1}{C^{4}}\sum\limits_{n=0}^{\infty }(n+1)\Gamma ^{2n}\rho
_{post}^{n}$. All the terms $\rho _{post}^{n}$ sum up incoherently.\ Let us
explain the latter procedure. In the actual conditions any $\left[ \left|
\psi _{-}^{n}\right\rangle \left\langle \psi _{-}^{n}\right| \right] $
leads, after the BS action, to two transmitted photons and $2(n-1)$
reflected photons. Different $n$ number of input pairs lead to the discard
of a different numbers of reflected photons, hence any mutual coherence is
destroyed after the tracing operation. The normalized density matrix turns
out to be: 
\begin{equation}
\rho _{th}^{II}=\left( 
\begin{array}{cccc}
\frac{1-p}{4} & 0 & 0 & 0 \\ 
0 & \frac{1+p}{4} & -\frac{p}{2} & 0 \\ 
0 & -\frac{p}{2} & \frac{1+p}{4} & 0 \\ 
0 & 0 & 0 & \frac{1-p}{4}
\end{array}
\right)  \label{Wernerstate}
\end{equation}

The SPDC density matrix $\rho _{th}^{II}$, given by the sum of Werner
states, is a Werner state itself, with singlet weight 
\begin{equation}
p=\frac{1}{2\widetilde{\Gamma }^{2}+1}  \label{Weight}
\end{equation}
with $\widetilde{\Gamma }=(1-\eta )\tanh g$. In the limit $\eta \rightarrow
0 $, $\widetilde{\Gamma }=\tanh g$. For large values of $g$\ , i.e. for $%
\widetilde{\Gamma }\rightarrow 1$, and in the hypothesis of very high
losses, the singlet weight $p\geq \frac{1}{3}$ approaches the minimum value $%
\frac{1}{3}$. Since the condition $p>\frac{1}{3}$ implies the well known
non-separability condition for a general Werner state, we have demonstrated
for large $g$ the expected high resilience against de-coherence of the
entangled singlet state \cite{Wern89}. The graph of Fig.2 shows the behavior
of singlet weight $p$ as a function of the interaction parameter $g$.

\section{Experimental realization}

The previous theoretical results have been experimentally tested for
different values of the parametric NL-gain $g$: Fig.3. The main source was a 
$Ti:Sa$ $Coherent$ $MIRA$ mode-locked laser further amplified by a $Ti:Sa$
regenerative $Coherent$ $Rega$ $9000$ device $\left( A\right) $\ operating
with pulse duration $180fs.$ The amplifier could operate either at a
repetition rate $250kHz$ or $100kHz$ leading to an energy per pulse,
respectively, of $4\mu J$ and $8\mu J$. The output beam was frequency
doubled in a UV beam at $\lambda _{p}=397.5nm$ through a {\it Second
Harmonic Generation }$\left( SHG\right) $ process, achieved by focusing the
infrared beam into a $1mm$ thick BBO crystal $(\beta -barium-borate)$, cut
for type I phase matching, through a lens with focal length of $20cm$. The
nonlinear (NL) crystal was placed at $5cm$ from the beam waist in order to
avoid crystal damage and beam spatial distortion. The UV beam then excited a 
$SPDC$ process in a $L=1.5mm$ thick $BBO$ NL crystal slab: Fig. 3. The $%
SPDC- $generated photons with degenerate wavelengths (wl's) $\lambda
=2\lambda _{p}=795nm$ propagated along the ${\bf k}_{1}$ and ${\bf k}_{2}$
modes$.$ A $\lambda /2$ waveplate (wp) and a $\frac{L}{2}$ thick $BBO$ were
placed on each mode to ensure the accurate compensation of all residual
birefringence effects coming from the main $BBO$ crystal, cut for type II
phase matching \cite{Kwia94}. In each mode ${\bf k}_{i}$, an additional
glass plate $\left( G_{p}\right) $ ensured a tight balance between the two
polarization emission cones of the SPDC\ process. The balancement between
the two cones was achieved by suitable tilting of $G_{p}$ in order to vary
the ratio between the transmittivities for the $s-$ and $p-$polarized waves.
Calibrated neutral attenuation filters $(At)$ placed in modes ${\bf k}_{1}$
and ${\bf k}_{2}$ were adopted to assure the condition of high-losses and
hence single-photon detection regime. The polarization states analysis was
carried out through two $\pi -$analyzers ($T_{1}$ and $T_{2}$ in Fig.3) each
one consisting of a pair of $\lambda /4+\lambda /2$ optical waveplates, a
polarizing beam splitter $(PBS),$ a single mode fiber coupled detector $%
SPCM-AQR14-FC$ with an interferential filter of bandwidth $\Delta \lambda
=4.5nm$ placed in front of it$.$ The combination of the UV $\lambda /2$\ wp (%
$WP_{P}$) and $PBS_{P}$ allowed a fine tuning of the UV pump power exciting
the NL crystal.

In a first experiment we estimated the gain value $\left( g\right) $ of the
optical parametric process and the overall quantum efficiencies of the
detection apparatus on both modes. The count rates of $D_{1}$ and $D_{2}$
and the coincidence rate $\left[ D_{1},D_{2}\right] $ were measured for
different values of the UV power (Fig.4). The plots of Figure 4-(a) and
4-(b) clearly show the onset of the NL parametric interaction with large $g$%
,thus implying the generation of many photon pairs. The gain value of the
process is obtained by fitting the count rates $N_{i}$ of detector $D_{i},$
dependent of the UV pump power, with the function $N_{i}(g)=R\frac{\eta
_{i}\Gamma ^{2}}{1-(1-\eta _{i})\Gamma ^{2}}$. Here $\eta _{i}$ is the
quantum efficiency on mode ${\bf k}_{i}$ and $R$ is the repetition rate of
the pump source \cite{Eise04}. The maximal value of gain obtained has been
found $g_{\max }=\left( 1.313\pm 0.002\right) ,$ which leads to a mean
photon number per mode $\overline{n}=\sinh ^{2}g_{\max }=\left( 2.97\pm
0.01\right) $. In conclusion the maximal total number of generated photon on 
${\bf k}_{1}$ and ${\bf k}_{2}$ modes through the SPDC\ process is $M=4%
\overline{n}=\left( 11.89\pm 0.05\right) $. By means of the previous fits we
could also estimated the overall detection efficiencies $\left( \eta
_{i}\right) $ on the ${\bf k}_{1}$ and ${\bf k}_{2}$ modes which depend on
the glass attenuation, the fiber coupling, the detection quantum
efficiencies: $\eta _{1}=(0.016\pm 0.002)$ and $\eta _{2}=(0.014\pm 0.002)$.
By the previous values we find $\eta \overline{n}\simeq 0.05$.

The main experimental result of the present work is the full
characterization of the 2-photon state.\ We reconstructed the density matrix 
$\rho _{\exp }^{II}$ of the generated $2-qubit$ state on ${\bf k}_{1}$ and $%
{\bf k}_{2}$ modes by adopting the Quantum State Tomography method $\left(
QST\right) $ \cite{Jame01}. The experimental density matrix$\ \rho _{\exp
}^{II}$ is obtained by first measuring the $2-$photon coincidences $\left[
D_{1},D_{2}\right] $ for different settings of the $QST$ setup, $T_{1}$ and $%
T_{2},$ and then by applying a numerical algorithm to estimate the density
matrix. In a low gain condition the SPDC\ state generated on ${\bf k}_{1}$
and ${\bf k}_{2}$ modes is expected to be in the singlet state $\left| \Psi
^{-}\right\rangle =2^{-1/2}\left( \left| H\right\rangle _{k_{1}}\left|
V\right\rangle _{k_{2}}-\left| V\right\rangle _{k_{1}}\left| H\right\rangle
_{k_{2}}\right) $, with excellent agreement between theory and experiment:
Figure 5-{\bf (d).} By increasing $g$, the $\rho $ elements corresponding to 
$\left| H\right\rangle _{k_{1}}\left| H\right\rangle _{k_{2}\text{ }%
k_{1}}\left\langle H\right| _{k_{2}}\left\langle H\right| $ and $\left|
V\right\rangle _{k_{1}}\left| V\right\rangle _{k_{2}\text{ }%
k_{1}}\left\langle V\right| _{k_{2}}\left\langle V\right| $ are no longer
negligible and the detection of two photon with same polarization is a
consequence of a multipairs condition: (Figure 5-{\bf (c), (b), (a))}. The
experimental results of the density matrices $\rho _{\exp }^{II}$ for
different $g-$values are in good agreement with the theoretical prediction $%
\rho _{th}^{II}$; the mean value of fidelity between the four comparison is $%
{\cal F}=\left( 0.996\pm 0.002\right) ,$ where ${\cal F}\left( \rho
_{th}^{II},\rho _{\exp }^{II}\right) =Tr^{2}\sqrt{\sqrt{\rho _{th}^{II}}\rho
_{\exp }^{II}\sqrt{\rho _{th}^{II}}}.$

The density matrices $\rho _{\exp }^{II}$ can now be adopted to estimate{\it %
\ ''singlet weight'', ''tangle'' }and ''{\it linear} {\it entropy''} of the
generated state. The density matrix $\rho _{W}$ of a Werner state is given
by the expression (\ref{Wernerstate}), as said. The singlet weight $\left(
p\right) $ can be directly obtained by the matrix elements as $p=\left( \rho
_{\exp }^{II}\right) _{22}+\left( \rho _{\exp }^{II}\right) _{33}-\left(
\rho _{\exp }^{II}\right) _{11}-\left( \rho _{\exp }^{II}\right) _{44}.$
Werner states are entangled $(p>\frac{1}{3})$ or separable $(p\leq \frac{1}{3%
})$, the extreme conditions being the pure singlet $\left( p=1\right) $ and
the totally mixed state $\left( p=0\right) $. The tangle is a parameter
expressing the degree of entanglement of the state, which is defined as $%
\tau =C^{2}$, where $C$ is the concurrence of the state \cite{Wei03}; $\tau
>0$ is a necessary and sufficient condition for a $2\times 2$ state to be
entangled. Another important property for a mixed state is {\it linear
entropy} $\left( S\right) ,$ which quantifies the degree of disorder, viz.
the mixedeness of the system. For a system of dimension $4${\it ,}\ it
results $S=\frac{4}{3}\left( 1-Tr\left( \rho ^{2}\right) \right) $. In case
of a Werner state, we have $S_{W}=\left( 1-p^{2}\right) $. For Werner
states, ''tangle'' and ''linear entropy'' are found to be related as follows 
\cite{Barb04}: 
\begin{equation}
\tau \left( S_{W}\right) =\left\{ 
\begin{array}{c}
\frac{1}{4}\left( 1-3\sqrt{1-S_{W}}\right) ^{2}\text{ \ \ for }0\leq
S_{W}\leq \frac{8}{9} \\ 
0\text{ \ \ \ \ \ \ \ \ \ \ \ \ \ \ \ \ \ \ \ \ \ \ \ \ \ \ \ \ \ \ \ for }%
\frac{8}{9}\leq S_{W}\leq 1
\end{array}
\right.  \label{Tangle}
\end{equation}
For each experimental value of $g$, $\left( S,\tau \right) $ are estimated
starting from the experimental density matrix. The agreement between
experimental results and theoretical predictions are found satisfactory:
Figure 6-{\bf (a)}.

An alternative method to establish whether a state is entangled or not is
based on the concept of entanglement witness. A state $\rho $ is entangled
if and only if there exists a Hermitian operator $\widehat{O}$, a so-called 
{\it entanglement witness}, which has positive expectation value $Tr\left[ 
\widehat{O}\rho _{sep}\right] \geq 0$ for all separable states $\rho _{sep}$
and has negative expectation value $Tr\left[ \widehat{O}\rho \right] <0$ on
the state $\rho $ \cite{Pere96,Terh01,Lewe00,Lewe01}. For Werner states $%
\rho _{W}$ the method proposed in \cite{Guhn02,DAri03} leads to the
following entanglement-witness operator: 
\begin{equation}
\widehat{O}_{W}=\frac{1}{2}\left( 
\begin{array}{c}
\left| H\right\rangle \left| H\right\rangle \left\langle H\right|
\left\langle H\right| +\left| V\right\rangle \left| V\right\rangle
\left\langle V\right| \left\langle V\right| +\left| D\right\rangle \left|
D\right\rangle \left\langle D\right| \left\langle D\right| \\ 
+\left| F\right\rangle \left| F\right\rangle \left\langle F\right|
\left\langle F\right| -\left| L\right\rangle \left| R\right\rangle
\left\langle L\right| \left\langle R\right| -\left| R\right\rangle \left|
L\right\rangle \left\langle R\right| \left\langle L\right|
\end{array}
\right)  \label{WitnessOperator}
\end{equation}
where $\left| D\right\rangle =\frac{1}{\sqrt{2}}\left( \left| H\right\rangle
+\left| V\right\rangle \right) $ and $\left| F\right\rangle =\frac{1}{\sqrt{2%
}}\left( \left| H\right\rangle -\left| V\right\rangle \right) $ express
diagonally polarized single photon states, while $\left| L\right\rangle =%
\frac{1}{\sqrt{2}}\left( \left| H\right\rangle +i\left| V\right\rangle
\right) $ and $\left| R\right\rangle =\frac{1}{\sqrt{2}}\left( \left|
H\right\rangle -i\left| V\right\rangle \right) $ express left and right
circular polarization states. The relationship between the expectation value
for a Werner state $W_{W}=Tr\left[ \widehat{O}_{W}\rho _{W}\right] $ and the
Werner weight $p$ is found to be 
\begin{equation}
W_{W}(p)=\frac{1-3p}{4}  \label{Witnesstheory}
\end{equation}
\cite{Cine04}, leading to $W_{W}(p)<0$ for $p>\frac{1}{3}$. Experimentally $%
Tr\left[ \widehat{O}_{W}\rho _{\exp }^{II}\right] $ has been estimated
through $8$ projective measurements (the $6$ projectors appearing in (\ref
{WitnessOperator}) and the operators $\left\{ \left| H\right\rangle \left|
V\right\rangle \left\langle H\right| \left\langle V\right| ,\left|
V\right\rangle \left| H\right\rangle \left\langle V\right| \left\langle
H\right| \right\} $ for normalization.\cite{Barb03}){\bf . }In conclusion,
for each $g$ value, a point of the Cartesian plane of coordinates $\left(
p,W\right) $ is obtained: Figure 6-{\bf (b)}. The solid line reports the
theoretical dependence (\ref{Witnesstheory}). The comparison demonstrates a
good agreement between the theoretical prediction and experimental results.

By the different methods described above the entanglement condition has been
found to be realized for a value of $g$ up to $1.084\pm 0.002$ (Fig.5-({\bf c%
})), corresponding to an average number of photons equal to $M=4\overline{n}%
=\left( 6.85\pm 0.03\right) $ before losses. For higher values of $g$ the
presence of bipartite entanglement is degraded by decoherence effects,
mostly due to imperfect correction of the walk-off effect in the BBO crystal
and to time distinguishability introduced by the femtosecond pump pulse.

\section{Conclusions}

In summary, the present work shows that the multiphoton states generated by
SPDC\ exhibit a bipartite entanglement even in the presence of high losses,
confirming previous analysis \cite{Durk04}. An explicit form has been
derived for the output two photon state: a Werner state. The theoretical
result are found to be in very good agreement with experimental data. We
believe that the present results could be useful to investigate the
resilience of entanglement in lossy communication. The present approach can
be extended to investigate the {\it quantum injected optical parametric
amplifier }${\it (QIOPA)}$ \cite{DeMa02,Pell03,DeMa04}.

We thank Marco Barbieri for useful discussions. Work\ supported by the FET
EU Network on QI Communication (IST-2000-29681: ATESIT), INFM (PRA\
''CLON'')\ and by MIUR (COFIN 2002).

Figure 1. Schematic layout. Inset:\ losses simulation by beamsplitter.

Figure 2. Werner weight $p$ versus non-linear parametric gain $g$. The
two-photon density matrices $\rho _{th}^{II}$ are reported for some gain
values ($g=0.1,g=1,g=0.3$).

Figure 3. Experimental setup adopted for multiphoton states generation by
means of $SPDC$ process and characterization by $QST$.

Figure 4. ({\bf a}) Count rates $[D_{1}]$ as a function of the UV power
(arbitrary unit). The continuous line expresses the best fit result. ({\bf b}%
) Coincidence rates $[D_{1},D_{2}]$ as a function of the UV power.

Figure 5. Theoretical $\rho _{th}^{II}$ (left plot) and experimental $\rho
_{\exp }^{II}$ (right plot) density matrices for different gain values. The
experimental density matrices have been reconstructed by measuring $16$ two
qubits observables through the two tomographic setups $\{T_{i}$\}$.$ Each
tomographic measurement lasted a time $t$ and yielded a maximum twofold
counts $\left( cc\right) $ for the $\left| HV\right\rangle $ projection (%
{\bf a}) $t=1\sec $; $cc\simeq 9300$ ({\bf b}) $t=2\sec $; $cc\simeq 12000$ (%
{\bf c}) $t=15\sec $; $cc\simeq 2000$ ({\bf d}) $t=120\sec $; $cc\simeq 1300$%
.

Figure 6.({\bf a}) The tangle parameter $\tau $ in function of the Entropy $%
S $ of the state. Red line; theoretical plot (\ref{Tangle}). ({\bf b})
Witness parameter $W=Tr\left[ \widehat{O}_{W}\rho _{\exp }^{II}\right] $ in
function of singlet weight $p.$ Red line: theoretical plot (\ref
{Witnesstheory}).


\begin{references}
\bibitem{Raus01}  R. Raussendorf , and H. J Briegel, Phys. Rev. Lett. {\bf 86%
}, 5188 (2001); P. Walther, {\it et al., }Nature (London) {\bf 434}, 169
(2005).

\bibitem{Boto00}  A.N. Boto, {\it et al.}, Phys. Rev. Lett. {\bf 85}, 2733
(2000).

\bibitem{Bosc97}  D. Boschi, S. Branca, F. De Martini, L. Hardy, and S.
Popescu, Phys. Rev. Lett. {\bf 80}, 1121 (1998); D. Bouwmeester, {\it et al.}%
, Nature (London) {\bf 390}, 575 (1997); E. Lombardi, F. Sciarrino, S.
Popescu and F. De Martini, Phys. Rev. Lett. {\bf 88}, 070402 (2002).

\bibitem{Gisi02}  N. Gisin, G. Ribordy, W. Tittel., H.\ Zbinden, Rev. Mod.
Phys. {\bf 74}, 145 (2002).

\bibitem{Kwia94}  P.G. Kwiat, K. Mattle, H. Weinfurter, and A. Zeilinger,
Phys. Rev. Lett.\ {\bf 75}, 4337 (1995).

\bibitem{Kwia99}  P G. Kwiat, E,\ Waks, A.G. White, I. Appelbaum, and P.H.
Eberhard, Phys. Rev. A\ {\bf 60}, R867 (1999).

\bibitem{Cine04}  C. Cinelli, G. Di Nepi, F. De Martini, M. Barbieri, and P.
Mataloni, Phys. Rev. A {\bf 70}, 022321 (2004).

\bibitem{Fior04}  M. Fiorentino {\it et al.}, Phys. Rev. A {\bf 69}, 041801
(2004).

\bibitem{Kurt01}  C. Kurtsiefer, M. Oberparleiter, and H. Weinfurter, Phys.
Rev. A {\bf 64,} 023802 (2001).

\bibitem{Durk03}  G.A. Durkin, C.Simon, and D. Bouwmeester, Phys. Rev. Lett. 
{\bf 88}, 187902 (2002).

\bibitem{DeMa2000}  F. De Martini {\it et al.} Optics Comm. 179, 581 (2000);
F. De Martini, {\it et al.}, Phys. Rev. Lett. {\bf 8}7, 150401 (2001).

\bibitem{Lama01}  A. Lamas-Linares, J.C. Howell, and D. Bouwmeester, Nature
(London) {\bf 412}, 887 (2001); J.C. Howell, A. Lamas-Linares, and D.
Bouwmeester, Phys. Rev. Lett. {\bf 88}, 030401 (2002).

\bibitem{Pell03}  D. Pelliccia, V. Schettini, F. Sciarrino, C. Sias, and F.
De Martini, Phys. Rev. A{\it \ }{\bf 68,} 042306 (2003).

\bibitem{DeMa04}  F. De Martini, D. Pelliccia, and F. Sciarrino, Phys. Rev.
Lett. {\bf 92}, 067901 (2004).

\bibitem{Eibl03}  M. Eibl, S. Gaertner, M. Bourennane, C. Kurtsiefer, M.
Zukowski, and H. Weinfurter, Phys. Rev. Lett. {\bf 90}, 200403 (2003).

\bibitem{Zhao04}  Z. Zhao, Y.-A. Chen, A.-N. Zhang, T. Yang, H.J. Briegel,
and J.-W. Pan, Nature (London) {\bf 430}, 54 (2004).

\bibitem{Dur00}  W. D\"{u}r, G. Vidal, and J. I. Cirac, Phys. Rev. A {\bf 62}%
, 062314 (2000).

\bibitem{Acin01}  A. Ac\'{i}n, D. Bru\ss , M. Lewenstein, and A. Sanpera,
Phys. Rev. Lett. {\bf 87}, 040401 (2001).

\bibitem{Eibl04}  M. Eibl, N. Kiesel, M. Bourennane, C. Kurtsiefer, and H.
Weinfurter, Phys. Rev. Lett. {\bf 92}, 077901 (2004).

\bibitem{Walt05b}  P. Walther, K.J. Resch, and A.\ Zeilinger, Phys. Rev.
Lett. {\bf 94}, 240501 (2005).

\bibitem{Reid02}  M.D. Reid, {\it et al.}, Phys. Rev. A {\bf 66}, 033801
(2002).

\bibitem{Simo03}  C. Simon and D. Bouwmeester, Phys. Rev. Lett. {\bf 91},
053601 (2003).

\bibitem{Eise04}  H. S. Eisenberg, G. Khoury, G. A. Durkin, C. Simon, and D.
Bouwmeester, Phys. Rev. Lett.{\it \ }{\bf 93,} 193901 (2004).

\bibitem{Wern89}  R. F. Werner, Phys Rev. A {\bf 40}, 4277 (1989).

\bibitem{Guhn02}  O. G\"{u}hne, {\it et al.}, Phys. Rev. A{\it \ }{\bf 66,}
062305 (2002).

\bibitem{Barb03}  M. Barbieri, F. De Martini, G. Di Nepi, and P. Mataloni,
Phys. Rev. Lett.{\it \ }{\bf 91,} 227901 (2003).

\bibitem{Durk04}  G.A. Durkin, C. Simon, J. Eisert, and D. Bouwmeester,
Phys. Rev. A{\it \ }{\bf 70,} 062305 (2004).

\bibitem{Vedr98}  V. Vedral and M. B. Plenio, Phys. Rev. A {\bf 57}, 1619
(1998).

\bibitem{DeMa05}  F.\ De Martini and F. Sciarrino, Progress in Quantum
Electronics {\bf 29}, 165 (2005).

\bibitem{Coll88}  M.J. Collett, Phys. Rev. A {\bf 38}, 2233 (1988).

\bibitem{Loud}  R. Loudon, {\it The Quantum Theory of Light, }3rd edition,
Oxford University Press, New York, 2000, par. 5.7, 6.10.

\bibitem{Niel00}  M. A. Nielsen and I. L. Chuang, {\it Quantum Computation
and Quantum Information} (Cambridge University, Cambridge, 2000).

\bibitem{Jame01}  D.F. V. James, P.G. Kwiat, W.J. Munro, and A.G. White,
Phys. Rev. A {\bf 64}, 052312 (2001)

\bibitem{Wei03}  T. Wei, {\it et al.}, Phys. Rev. A {\bf 67}, 022110 (2003).

\bibitem{Barb04}  M. Barbieri, F. De Martini, G. Di Nepi, and P. Mataloni,
Phys. Rev. Lett. {\bf 92}, 177901 (2004).

\bibitem{Pere96}  A. Peres, Phys. Rev. Lett. {\bf 77, }1413 (1996).

\bibitem{Terh01}  B. Terhal, Lin. Alg. Appl. {\bf 323}, 61 (2001).

\bibitem{Lewe00}  M. Lewenstein, B. Kraus, J. I. Cirac, and P. Horodecki,
Phys. Rev. A {\bf 62}, 052310 (2000).

\bibitem{Lewe01}  M. Lewenstein, B. Kraus, P. Horodecki, and J. I. Cirac,
Phys. Rev. A {\bf 63}, 044304 (2001).

\bibitem{Brub02}  D. Bru\ss , J. I. Cirac, P. Horodecki, F. Hulpke, B.
Kraus, M. Lewenstein, and A. Sanpera, J. Mod. Opt. {\bf 49}, 1399 (2002).

\bibitem{DAri03}  G.M. D'Ariano, C. Macchiavello and M.G.A. Paris, Phys.
Rev. A {\bf 67}, 042310 (2003).

\bibitem{DeMa02}  F. De Martini, V. Bu\v{z}ek, F. Sciarrino, and C. Sias,
Nature (London)\ {\bf 419}, 815 (2002).
\end{references}
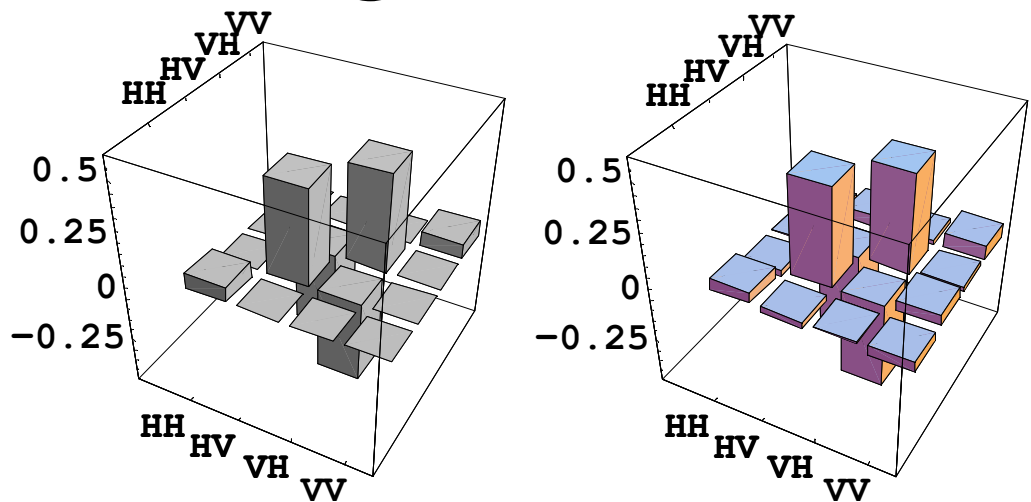
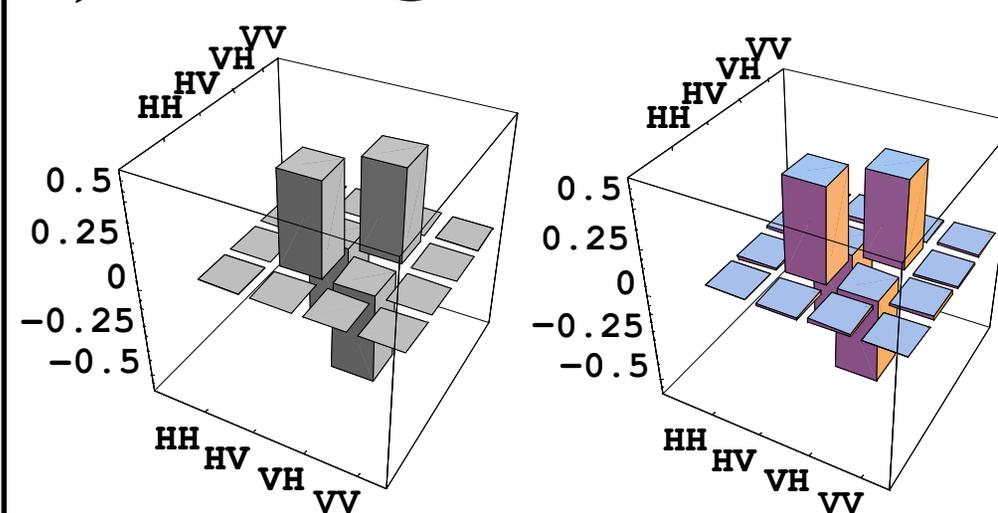
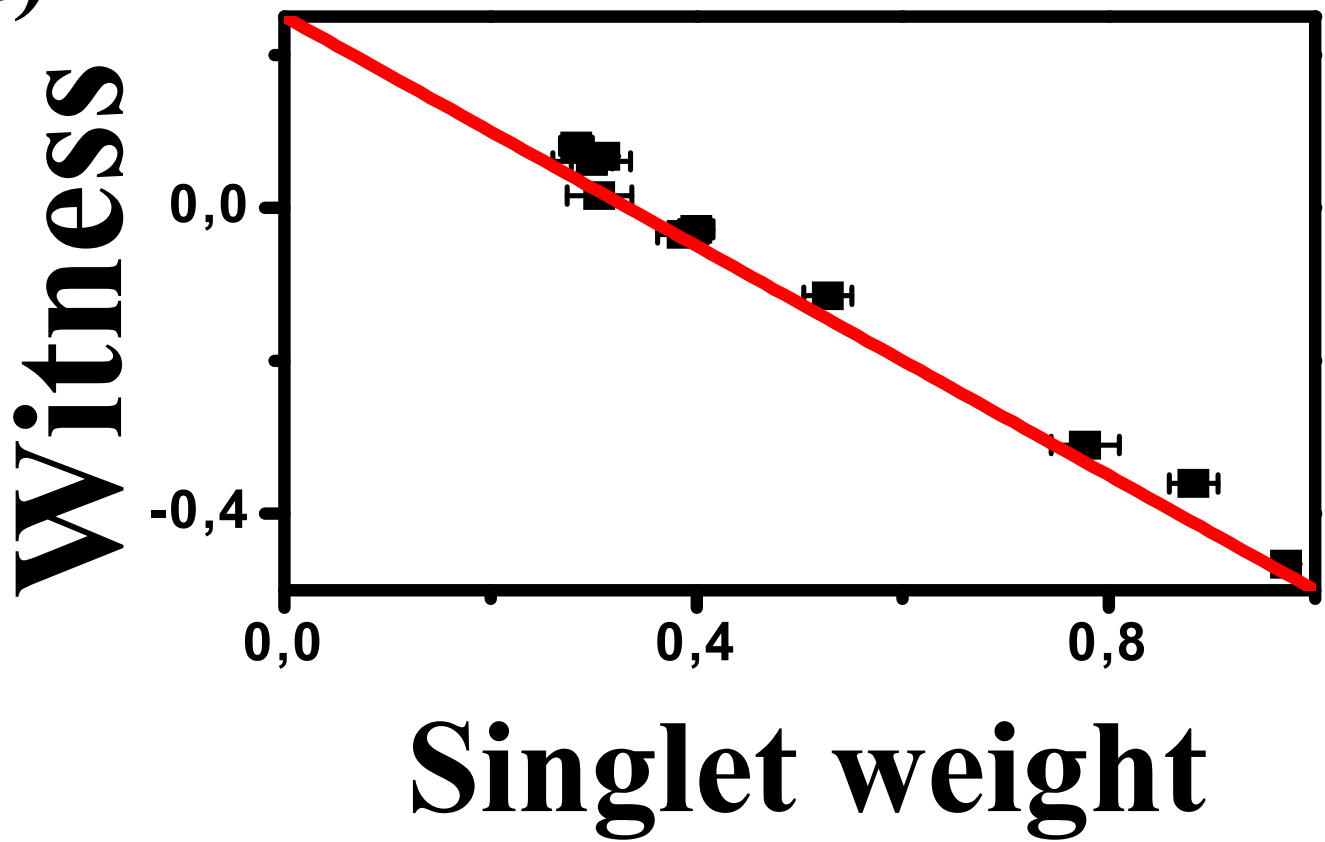
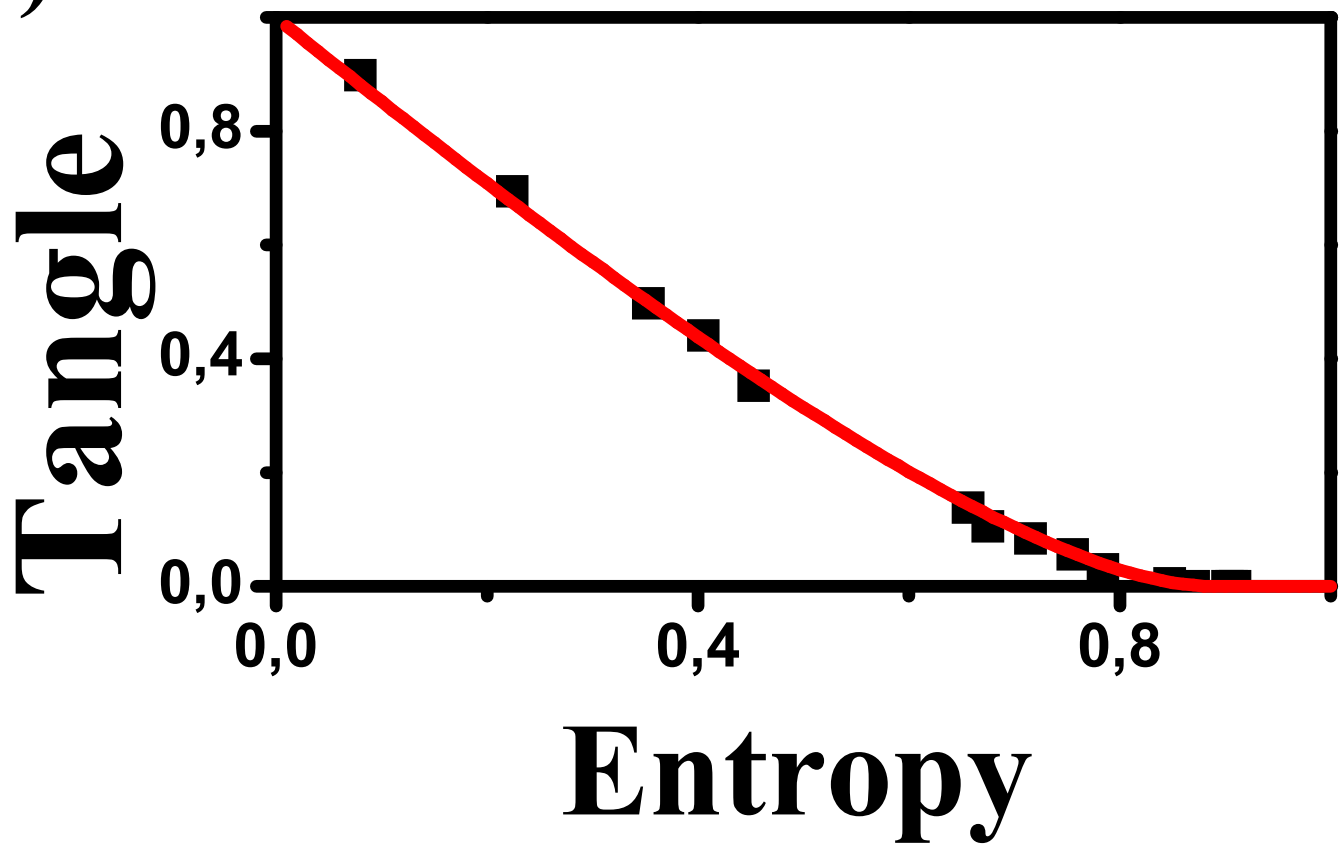
\end{document}